\pdfoutput=1
\documentclass[a4paper,fleqn,usenatbib]{mnras}
\usepackage{graphicx}	
\usepackage{amsmath}	
\usepackage{amssymb}	
\usepackage{bm}
\usepackage{txfonts}

\usepackage[T1]{fontenc}
\usepackage{ae,aecompl}
\usepackage{hyperref}
\usepackage{color}

\title[Imprint of PGW on gravitational lens system]{Imprint of primordial gravitational wave with extremely low frequency on gravitational lens system}

\author[Wenshuai Liu]{
Wenshuai Liu\thanks{E-mail: 674602871@qq.com (DPF)}\\
School of Physics, Henan Normal University, Xinxiang 453007, China\\
}

\date{Accepted XXX. Received YYY; in original form ZZZ}

\begin{document}
\label{firstpage}
\pagerange{\pageref{firstpage}--\pageref{lastpage}}
\maketitle

\begin{abstract}
Primordial gravitational waves with extremely low frequency are expected to origin from inflation in the early Universe. The detection of such kind of gravitational waves is of great significance to verify the inflationary theory and determine the energy scale of inflation. B-mode of polarization of cosmic microwave background shows to be a promising probe of extremely low frequency primordial gravitational wave. In order to find a complementary observational feature induced by primordial gravitational wave with extremely low frequency, we propose an alternative way of detection by investigating the effect of primordial gravitational wave with extremely low frequency on a gravitational lens system with a non-aligned source-deflector-observer configuration. The results show that, with a series of chosen parameters, gravitational lens system with perturbation from extremely low frequency primordial gravitational wave could induce time delay which could deviate from the time delay deduced from theoretical model as much as about one hundred percent, meaning that gravitational lens system with a non-aligned configuration could serve as a potential long-base-line detector of extremely low frequency primordial gravitational wave.
\end{abstract}

\begin{keywords}
gravitational lensing: strong -- gravitational waves -- methods: analytical
\end{keywords}



\section{Introduction}

Inflation, according to which a period of accelerated expansion occurs in the very early Universe, is the leading theory as it provides the seeds giving rise to structure formation and leads to a flat, homogeneous and isotropic Universe \citep{1,2,3,4}. Besides the scalar perturbations, inflation also produces tensor perturbations - primordial gravitational waves (PGWs) with a nearly scale-invariant spectrum through quantum fluctuations of the metric tensor \citep{5,6,7,8,9,10,11}. The detection of PGWs is of great significance since it would verify the inflationary scenario and determine the energy scale of inflation. The B-modes of polarization of cosmic microwave background (CMB) induced by extremely low frequency PGWs via Thomson scattering shows to be a promising way to detect PGW with frequency in the range of $10^{-18}$Hz-$10^{-16}$Hz \citep{12,13}.

Although B-mode of polarization of CMB is a promising probe of extremely low frequency PGW, polarized emission from dust in our Milky Way galaxy \citep{14} provides contamination, thus, it is worth investigating complementary observational techniques to confirm PGWs besides the method using B-mode of polarization of CMB. Based on the idea that very long-wavelength perturbations on the scale of galaxy formation act as an effective tidal field leading to non-zero E and B modes on large scales and altering cosmic shear and galaxy clustering spectrum measurements, a growing body of work \citep{15,16,17,18,19,20,21,22,23,24,25,26,27} has paid attention to the effect of extremely low frequency PGW on large-scale structure. However, such proposal also faces challenges mainly from gravitational nonlinearities which can source both E and B modes at relatively large scales and the stochasticity induced by small-scale perturbations. \citet{28} investigated the 21 cm fluctuations lensed by PGW to infer the amplitude of PGWs. \citet{29,30} studied the way of detecting PGWs by using the cross correlation of the galaxy distribution with Sunyaev-Zel'dovich effect.

Detection of extremely low frequency PGW based on the idea of using gravitational lens is first proposed by \citet{31,32} which suggested that gravitational lens could be used as the detector of such kind of PGWs with the additional time delays between different images of a quasar induced by perturbation from extremely low frequency PGW. Later on, work in \citet{33}, with the conclusion that time delays induced by PGW cannot be observationally distinguishable from the intrinsic time delays resulting from the geometry of the gravitational lens system, showed that gravitational lens could't act as a detector to probe PGW with extremely low frequency by using time delays between different images of a quasar.

Recently, a new method to detect extremely low frequency PGWs using gravitational lens is proposed by \citet{34} where time delay with a special relationship shown in Eq. (26) in \citet{34} is the evidence of PGW with extremely low frequency traveling through the lens system with the condition that the source and the observer are equidistant from and aligned with the deflector and that $\omega L \eta \ll 1$ and $h \ll \eta$ where $\eta$, $\omega$, $h$ and $L$ are the Einstein radius, the frequency and the dimensionless amplitude of the gravitational wave and the distance from the source (or the observer) to the deflector, respectively. Results from \citet{35} show Eq. (26) in \citet{34} holds in the general condition that the distance from the observe to the deflector is not equal to that from the source to the deflector, the direction of propagation and polarization of the gravitational wave are arbitrary, and the linear superposition of PGWs is present, meaning that gravitational lens with an aligned source-deflector-observer configuration could serve as a detector of extremely low frequency PGW when the whole Einstein ring is taken into account.

However, the possibility of having gravitational lens with a highly aligned source-detector-observer configuration is extremely low in the Universe and lens systems with a non-aligned configuration are common. In order to investigate whether the non-aligned gravitational lens system could be used as a long-base-line detector of extremely low frequency PGW, effect of PGWs with extremely low frequency on gravitational lens with a non-aligned configuration is investigated in this work. The trajectory of light in the non-aligned gravitational lens system perturbed by extremely low frequency PGW is derived in detail and the resulting time delay between different images of the source is obtained quantitatively. We find that extremely low frequency PGW could induce time delay which could present obvious deviation from the time delay deduced from theoretical model, meaning that the non-aligned gravitational lens system could act as a promising long-base-line detector of extremely low frequency PGW. We present this new method in Section 2 in detail. Conclusions are in Section 3.

\section{Perturbations of PGW on gravitational lens system}
\begin{figure}
\begin{center}
\includegraphics[clip,width=0.5\textwidth]{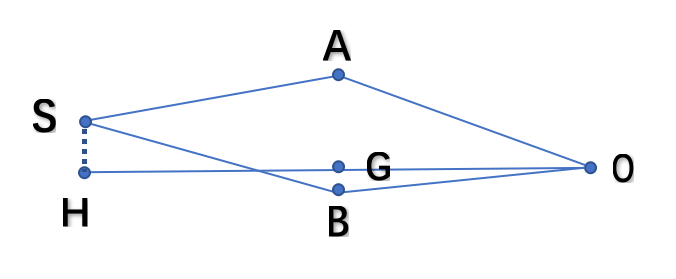}
\caption{ $S$, $G$ and $O$ represent the source, the deflector and the observer, respectively. The lens axis is the z axis. $G$ and $O$ are on the z axis with coordinates $G$ $(x=0,y=0,z=0)$ and $O$ $(x=0,y=0,z=L)$. $A$ with coordinates $(x=R_1,y=0,z=0)$ and $B$ with coordinates $(x=R_2,y=0,z=0)$ are the points of deflection due to gravitation of $G$. The line consisted of $A$ and $G$ and the line consisted of $B$ and $G$ are perpendicular to the line consisted of $O$ and $G$. The angle between the line consisted of $S$ and $O$ and the line consisted of $G$ and $O$ is $\beta$. The line consisted of $S$ and $H$ with coordinates $(x=0,y=0,z=-L)$ is perpendicular to the line consisted of $O$ and $G$.}
\end{center}
\end{figure}

The lens geometry with a non-aligned source-deflector-observer configuration is shown in Figure 1 where the distance from the observer to the deflector is equal to that from the projection of the source on the line consisted of the observer and the deflector to the deflector. A point (or thin axially symmetric) gravitational deflector with mass $M$ is at the origin of coordinates. The source and the observer are at ($x=2L\beta,y=0,z=-L$) and ($x=0,y=0,z=L$), respectively. The gravitational potential of the deflector is $U$ and the extremely low frequency PGW propagates through the lens system where we assume that the direction of propagation with an inclination angle $\phi$ with respect to $z$ axis lies in the $x-z$ plane. The speed of light is set to be $c=1$. Thus, the metric of the gravitational wave is
\begin{eqnarray}
h_{ij}=&&
\left[\begin{array}{cccc}
    -\cos^2\phi\hspace{1 mm} h_+ &    -\cos\phi\hspace{1 mm} h_\times    & \sin\phi\hspace{1 mm}\cos\phi\hspace{1 mm} h_+ \\
    -\cos\phi\hspace{1 mm} h_\times &    h_+   & \sin\phi\hspace{1 mm} h_\times\\
    \sin\phi\hspace{1 mm}\cos\phi\hspace{1 mm} h_+ & \sin\phi\hspace{1 mm} h_\times & -\sin^2\phi\hspace{1 mm} h_+
\end{array}\right]
\nonumber \\
&& \times  \cos(\omega t-\mathbf{k} \cdot \mathbf{x})
\end{eqnarray}
where $\mathbf{k}=\omega(\sin\phi,0,\cos\phi)$ is the propagation vector, $\omega$ is the frequency, $h_+$ and $h_\times$ are the amplitude of the two polarizations of the gravitational wave, respectively.

Here, for simplicity, we set $\phi=\frac{\pi}{2}$ and $h_\times=0$. Then we get the total metric

\begin{eqnarray}
ds^2&=&(1+2U)dt^2-(1-2U)(dx^2+dy^2+dz^2)
\nonumber \\
&&+h_{ij}dx^idx^j\label{1}
\end{eqnarray}
where $U=-\frac{GM}{r}$.

The time of travel of light is \citep{33}
\begin{eqnarray}
T &\approx & \int_{-L}^{L}dz[1+\frac{1}{2}\left(\frac{dx}{dz}\right)^2+\frac{1}{2}\left(\frac{dy}{dz}\right)^2
\nonumber \\
&&+\frac{1}{2}h_{ij}\frac{dx^i}{dz}\frac{dx^j}{dz}-2U]\label{2}
\end{eqnarray}

The first three terms in the bracket of Eq. (\ref{2}) arise from the lens geometry and the fifth is due to the gravitational potential of the deflector. The fourth term is due to the contribution from gravitational wave. Here, $t$ in Eq. (\ref{2}) is replaced by $t=t_e+(z+L)$ in order to approach the level of approximation \citep{33} where $t_e$ is the time the photons were emitted at $(x=2L\beta,y=0,z=-L)$ and $\omega t_e$ acts as the initial phase.

With the method of Fermat's principle shown in \citet{33}, we derive the trajectory of light in gravitational lens system with perturbation from extremely low frequency PGW. The trajectory of light starts from point S ($x=2L\beta,y=0,z=-L$) and arrives at point O ($x=0,y=0,z=L$) after being deflected in the $z=0$ plane by the deflector G.

When $z<0$, the geodesic equation of light is written as follows \citep{33}

\begin{equation}
\frac{dx}{dz}=\gamma-\frac{1}{2}h\cos[\omega(t_e+z+L)]\label{3}
\end{equation}
where $\gamma$ is an integration constant which will be obtained below and $h=h_+$.

Integrating Eq. (\ref{3}) from $z=-L$ to $z=0$, we get the change in the direction of x-axis as

\begin{equation}
\Delta x_1=-\frac{1}{2}\frac{h}{\omega}\sin[\omega(t_e+L)]+L\gamma+\frac{1}{2}\frac{h}{\omega}\sin(\omega t_e)\label{4}
\end{equation}

Then we get

\begin{eqnarray}
x|_{z=0}&=&2L\beta+\Delta x_1
\nonumber \\
&=&2L\beta+L\gamma
\nonumber \\
&&+\frac{1}{2}\frac{h}{\omega}\sin(\omega t_e)-\frac{1}{2}\frac{h}{\omega}\sin[\omega(t_e+L)]\label{5}
\end{eqnarray}

When $z>0$, the geodesic equation of light is as follows after being deflected at point $A$ or $B$ with a deflected angle $\alpha=-\frac{4GM}{R}$ where $R=R_{1,2}=x_{A,B}=x|_{z=0}$

\begin{equation}
\frac{dx}{dz}=\gamma-\frac{4GM}{R}-\frac{1}{2}h\cos[\omega(t_e+z+L)]\label{6}
\end{equation}

Integrating Eq. (\ref{6}) from $z=0$ to $z=L$, we get the change in the direction of x-axis as

\begin{eqnarray}
\Delta x_2&=&L\gamma-\frac{4GM}{R}L
\nonumber \\
&&-\frac{1}{2}\frac{h}{\omega}\sin[\omega (t_e+2L)]+\frac{1}{2}\frac{h}{\omega}\sin[\omega(t_e+L)]\label{7}
\end{eqnarray}

Due to the fact that
\begin{equation}
x|_{z=0}=-\Delta x_2=R\label{8}
\end{equation}

we obtain $\gamma$ with Eq. (\ref{5}) and Eq. (\ref{7})
\begin{eqnarray}
\gamma &=& \frac{2GM}{R}+\frac{1}{4}\frac{h}{\omega L}\sin[\omega (t_e+2L)]
\nonumber \\
&&-\frac{1}{4}\frac{h}{\omega L}\sin(\omega t_e)-\beta\label{9}
\end{eqnarray}

Then we get

\begin{equation}
R = L\beta^*+\frac{2GM}{R}L\label{10}
\end{equation}

where
\begin{equation}
\beta^*=\beta+\beta_0\label{11}
\end{equation}

and

\begin{eqnarray}
\beta_0 &=& \frac{1}{4}\frac{h}{\omega L}\sin(\omega t_e)-\frac{1}{2}\frac{h}{\omega L}\sin[\omega(t_e+L)]
\nonumber \\
&&+\frac{1}{4}\frac{h}{\omega L}\sin[\omega(t_e+2L)]\label{12}
\end{eqnarray}

From Eq. (\ref{10}), we have
\begin{equation}
R_{1,2}=\frac{L\beta^*\pm\sqrt{L^2\beta^{*2}+8GML}}{2}\label{13}
\end{equation}

Then the slope at point $O$ is

\begin{equation}
\frac{dx}{dz}|_{z=L}=-\beta^*-\theta_0-\frac{2GM}{R_{1,2}}\label{14}
\end{equation}

where

\begin{eqnarray}
\theta_0&=&\frac{1}{2}h\cos[\omega(t_e+2L)]-\frac{1}{2}\frac{h}{\omega L}\sin[\omega(t_e+2L)]
\nonumber \\
&&+\frac{1}{2}\frac{h}{\omega L}\sin[\omega(t_e+L)]\label{15}
\end{eqnarray}

After transforming the slope to the angular position of the image of the source with respect to the line that joins the observer and the deflector, we get
\begin{eqnarray}
\theta_{1,2}&=&-\frac{dx}{dz}|_{z=L}
\nonumber \\
&=&\beta^*+\theta_0+\frac{2GM}{R_{1,2}}
\nonumber \\
&=&\frac{\beta^*}{2}\pm\frac{\sqrt{\beta^{*2}+4\eta^2}}{2}+\theta_0\label{16}
\end{eqnarray}
where $\eta=\sqrt{\frac{2GM}{L}}$.

With the same method adopted above, we get the angular position of the image of the deflector with respect to the line that joins the observer and the deflector
\begin{eqnarray}
\theta_d&=&\frac{1}{2}h\cos[\omega(t_e+2L)]-\frac{1}{2}\frac{h}{\omega L}\sin[\omega(t_e+2L)]
\nonumber \\
&&+\frac{1}{2}\frac{h}{\omega L}\sin[\omega(t_e+L)]\label{17}
\end{eqnarray}

It shows that $\theta_d=\theta_0$. Then, from the point of view of real observation, the image angular position of the source is the image of the source with respect to the image of the deflector and is expressed as

\begin{eqnarray}
\theta^{'}_{1,2}&=&\theta_{1,2}-\theta_d
\nonumber \\
&=&\frac{\beta^*}{2}\pm\frac{\sqrt{\beta^{*2}+4\eta^2}}{2}\label{18}
\end{eqnarray}

According to \citet{42}, parameters of $h=10^{-5}$, $\omega=10^{-18}$ are adopted. We set $\beta=0.7\times 10^{-6}$, $\eta=4.8\times 10^{-6}$ and $L=1Gpc$ in the following study. With Eq. (\ref{16}) and Eq. (\ref{18}), $\theta_{1,2}$ and $\theta^{'}_{1,2}$ are shown in Figure 2 where we find that the difference between $\theta_{1,2}$ and $\theta^{'}_{1,2}$ is minor with perturbations from extremely low frequency PGW.

\begin{figure}
\begin{center}
\includegraphics[clip,width=0.5\textwidth]{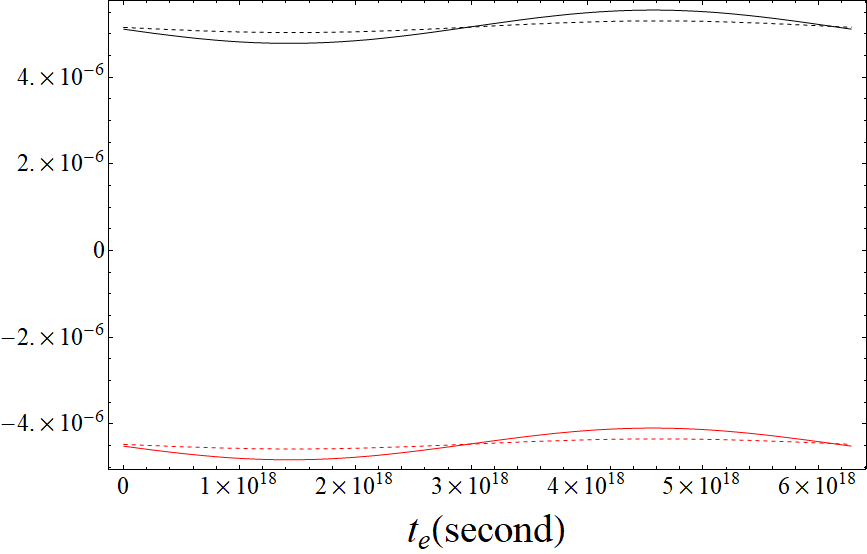}
\caption{Black and red lines represent $\theta_1$ and $\theta_2$ along with $t_e$, respectively. Black dotted and red dotted lines represent $\theta^{'}_1$ and $\theta^{'}_2$ along with $t_e$, respectively.}
\end{center}
\end{figure}

Based on $\theta_{1,2}$ in Eq. (\ref{16}), the actual magnification is expressed as $\mu=\frac{\theta}{\beta}\frac{d\theta}{d\beta}$ which could be shown as follows in detail
\begin{equation}
\mu=\frac{\theta(\theta-\theta_0)^3}{[(\theta-\theta_0)^2-\beta_0(\theta-\theta_0)-\eta^2][(\theta-\theta_0)^2+\eta^2]}
\end{equation}
while the magnification deduced from observation based on $\theta^{'}_{1,2}$ in Eq. (\ref{18}) is
\begin{equation}
\mu^{'}=\frac{\theta^{'}}{\beta^*}\frac{d\theta^{'}}{d\beta^*}=\frac{\theta^{'4}}{\theta^{'4}-\eta^4}
\end{equation}
where the misalignment angle $\beta^*$ is inferred by the observer through $\beta^*=\theta^{'}_1+\theta^{'}_2$ and $\eta$ through $\eta=\sqrt{-\theta^{'}_1\theta^{'}_2}$.
Then we get $\nu=\left|\frac{\mu_1}{\mu_2}\right|$ and $\nu^{'}=\left|\frac{\mu^{'}_1}{\mu^{'}_2}\right|$. The ratio $\frac{\nu}{\nu^{'}}$ is presented in Figure 3 where it shows that the effect of extremely low frequency PGW on magnification is not obvious.

\begin{figure}
\begin{center}
\includegraphics[clip,width=0.5\textwidth]{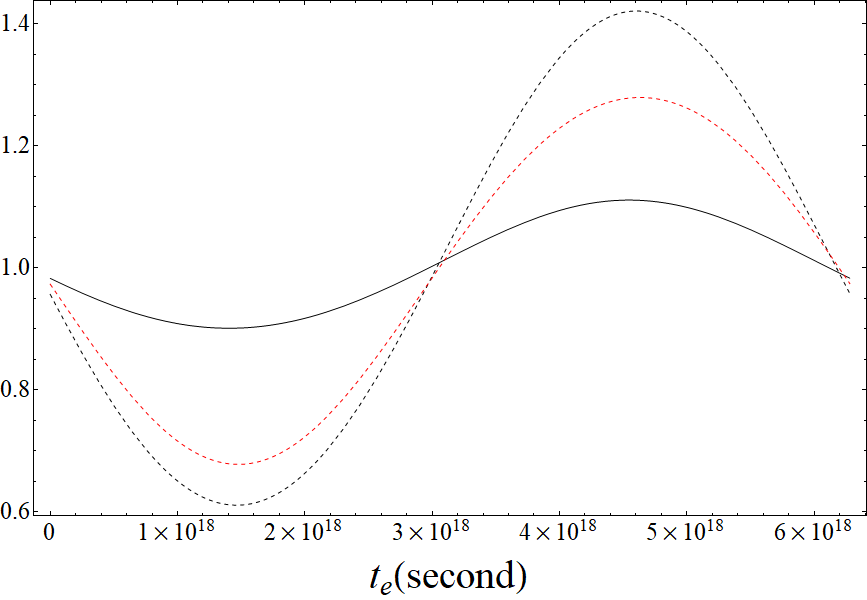}
\caption{Black, black dotted and red dotted lines represent $\frac{\nu}{\nu^{'}}$, $\frac{\mu_1}{\mu_1^{'}}$ and $\frac{\mu_2}{\mu_2^{'}}$, respectively.}
\end{center}
\end{figure}

From Eq. (\ref{2}), we give the time of travel of light as follows in detail
\begin{equation}
T \approx  \int_{-L}^{L}dz[1+\frac{1}{2}\left(\frac{dx}{dz}\right)^2
+\frac{1}{2}h_{33}\frac{dz}{dz}\frac{dz}{dz}+\frac{2GM}{\sqrt{x^2+z^2}}]\label{19}
\end{equation}

For actual trajectory of light, we have

\begin{equation}
\frac{dx}{dz}=\gamma-\frac{1}{2}h\cos[\omega(t_e+z+L)]\label{20}
\end{equation}

\begin{equation}
x=\gamma z-\frac{1}{2}\frac{h}{\omega}\sin[\omega(t_e+z+L)]+L\gamma+\frac{1}{2}\frac{h}{\omega}\sin(\omega t_e)+2L\beta\label{21}
\end{equation}
for $z<0$ and
\begin{equation}
\frac{dx}{dz}=\gamma-\frac{4GM}{R}-\frac{1}{2}h\cos[\omega(t_e+z+L)]\label{22}
\end{equation}

\begin{eqnarray}
x&=&\gamma z-\frac{4GM}{R}z
\nonumber \\
&&-\frac{1}{2}\frac{h}{\omega}\sin[\omega(t_e+z+L)]-L\gamma
\nonumber \\
&&+\frac{4GM}{R}L+\frac{1}{2}\frac{h}{\omega}\sin[\omega(t_e+2L)]\label{23}
\end{eqnarray}
for $z>0$.

For trajectory of light deduced from observation, it shows that
when $z<0$
\begin{equation}
\frac{dx}{dz}=\gamma^{'}\label{24}
\end{equation}

\begin{equation}
x=\gamma^{'} z+L\gamma^{'}+2L\beta^*\label{25}
\end{equation}
and when $z>0$
\begin{equation}
\frac{dx}{dz}=\gamma^{'}-\frac{4GM}{R}\label{26}
\end{equation}

\begin{equation}
x=\gamma^{'} z-\frac{4GM}{R}z-L\gamma^{'}+\frac{4GM}{R}L\label{27}
\end{equation}
where $\gamma^{'}=\frac{2GM}{R}-\beta^*$

\begin{figure}
\begin{center}
\includegraphics[clip,width=0.5\textwidth]{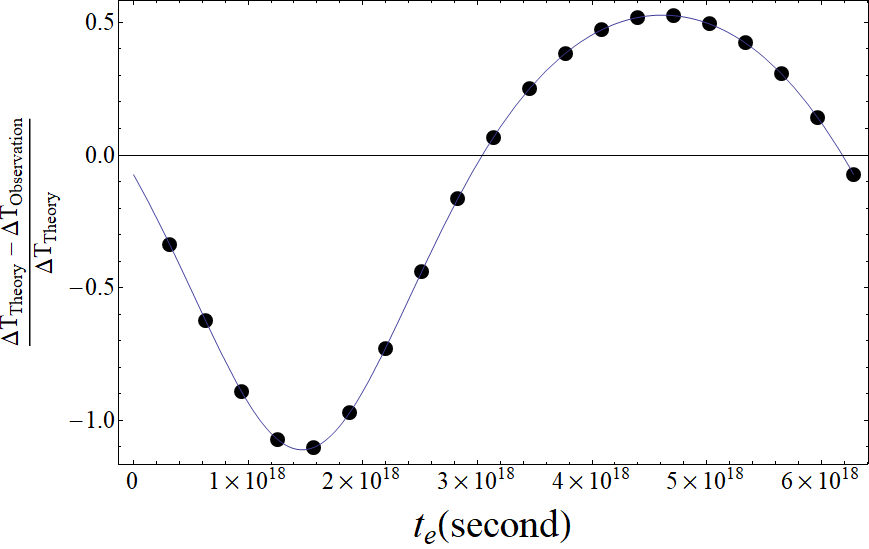}
\caption{Black dots represent $\frac{\Delta \rm T_{\rm{Theory}}-\Delta \rm T_{\rm {Observation}}}{\Delta \rm T_{\rm {Theory}}}$ along with $t_e$ using numerical calculation based on equations from Eq. (\ref{19}) to Eq. (\ref{27}) and the curve represents $\frac{\Delta \rm T_{\rm{Theory}}-\Delta \rm T_{\rm {Observation}}}{\Delta \rm T_{\rm {Theory}}}$ along with $t_e$ using Eq. (\ref{28}) and Eq. (\ref{29})}
\end{center}
\end{figure}

We define $\rm T_{\rm SAO}$/$\rm T_{\rm SBO}$ which represents the time travel of the light traveling from the source S through the point $\rm A$/$\rm B$ to the observer O. Based on equations from Eq. (\ref{19}) to Eq. (\ref{27}), we could derive the actual time delay between different images of the source, $\Delta \rm T_{\rm {Observation}}$ which can be expressed as
\begin{eqnarray}
\Delta  T_{Observation}&=&  T_{SAO}- T_{SBO}
\nonumber \\
&=&\frac{(2GM)^2}{L}\left(\frac{1}{\theta^{'2}_1}-\frac{1}{\theta^{'2}_2}\right)-4GM ln \left|\frac{\theta^{'}_1}{\theta^{'}_2}\right|+\frac{2GMh}{\omega}\left(\frac{1}{R_A}-\frac{1}{R_B}\right)
\nonumber \\
&&\times\{\sin(\omega t_e)-2\sin[\omega(t_e+L)]+\sin[\omega(t_e+2L)]\} \label{28}
\end{eqnarray}
and the time delay deduced from theoretical model, $\Delta \rm T_{\rm {Theory}}$ which is
\begin{equation}
\Delta  T_{Theory}=\frac{(2GM)^2}{L}\left(\frac{1}{\theta^{'2}_1}-\frac{1}{\theta^{'2}_2}\right)-4GM ln \left|\frac{\theta^{'}_1}{\theta^{'}_2}\right| \label{29}
\end{equation}

The value of $\frac{\Delta \rm T_{\rm {Theory}}-\Delta \rm T_{\rm {Observation}}}{\Delta \rm T_{\rm {Theory}}}$ is presented in Figure 4 where it shows that gravitational lens system with perturbation from extremely low frequency PGW could produce time delay which could deviate from the time delay inferred from theoretical model as much as about one hundred percent and that the analytic result is coincident with the numerical one.

In addition to time delay induced by PGW of extremely low frequency, other perturbations on time delay include a precise model of the deflector potential across the images to infer the relative Fermat potential (about a fraction of one day due to substructure) \citep{36} and the line-of-sight weak lensing effect (about a few percent or as much as ten percent) \citep{37,38,39} both of which could be inferred with independent data from high-resolution Hubble Space Telescope imaging of the lensing arc, the environment of the lens and the kinematics of the lensing galaxy \citep{40}. Furthermore, gravitational microlensing also produces additional time delay (order of days when the source in Figure 1 is a quasar) \citep{41}, such additional time delay could be eliminated by transient source such as repeated fast radio burst or gravitational wave with electromagnetic counterpart in binary neutron star merger. Compared with perturbations from extremely low frequency PGWs, other perturbations list above are weak. Thus, with the features of time delay shown in Figure 4, gravitational lens system with a non-aligned configuration shows to be a potential long-base-line detector of extremely low frequency PGW.
	

\section{CONCLUSIONS} \label{discuss}
In this work, we study the effect of extremely low frequency PGW on gravitational lens system with a non-aligned source-deflector-observer configuration where the projection of the source on the lens axis and the observer are equidistant from the deflector in order to find the possible evidence of the presence of PGW with extremely low frequency in the Universe. Results of this work show that, compared with magnification, time delay between different images of the source in the non-aligned lens configuration could be perturbed strongly by PGW of extremely low frequency and may show obvious deviation from time delay obtained from theoretical models in some situations. Thus, gravitational lens system shows to be a potential long-base-line detector of extremely low frequency PGW.

Due to the foreground contamination from dust in our Milky Way, detection of extremely low frequency primordial gravitational wave with the traditional method using B-mode polarization faces challenges. In order to investigate an alternative observational feature originating from such primordial gravitational wave, this work proposes a new way to detect extremely low frequency primordial gravitational wave by using time delay between different images in gravitational lensing. With large numbers of gravitational lens systems which will be discovered in future, there might be a chance to confirm the existence of extremely low frequency primordial gravitational wave with this new method.

Although we investigate the imprint of extremely low frequency PGW on a non-aligned lens configuration with the condition that the projection of the source on the lens axis and the observer are equidistant from the deflector, we may also expect that the same features hold with the general condition that distance from the observe to the defector is not equal to that from the projection of the source on the lens axis to the defector, the direction of propagation and polarization of PGWs are arbitrary, the deflector is not limited to a point (or thin axially symmetric) gravitational defector and the linear superposition of PGWs is present.

\textbf{Data Availability}

The data underlying this article will be shared on reasonable request to the corresponding author.



\bsp	
\label{lastpage}

\begin{thebibliography}{99}



\bibitem[Abbott \& Wise(1984)]{5}Abbott L. F., \& Wise M. B.\ 1984, Nucl. Phys., B244, 541

\bibitem[Ade et al.(2015)]{14}Ade P. A. R., et al., (BICEP2, Planck Collaborations)\ 2015, Phys. Rev. Lett., 114, 101301

\bibitem[Allen(1988)]{11}Allen B.\ 1988, Phys. Rev. D, 37, 2078

\bibitem[Allen(1989)]{31}Allen B.\ 1989, Phys. Rev. Lett., 63, 2017

\bibitem[Allen(1990)]{32}Allen B.\ 1990, Gen. Relativ. Gravit., 22, 1447

\bibitem[Alizadeh \& Hirata(2012)]{29}Alizadeh E., \& Hirata C.\ 2012, Phys. Rev. D, 85, 123540



\bibitem[Baskaran et al.(2006)]{42}Baskaran D., Grishchuk L. P., \& Polnarev A. G.\ 2006, Phys. Rev. D, 74, 083008








\bibitem[Bar-Kana(1996)]{38}Bar-Kana R.\ 1996, ApJ, 468, 17







\bibitem[Book et al.(2012)]{28}Book L., Kamionkowski M., \& Schmidt F.\ 2012, Phys. Rev. Lett., 108, 211301

\bibitem[Dai et al.(2013)]{22}Dai L., Jeong D., \& Kamionkowski M.\ 2013, Phys. Rev. D, 88, 043507

\bibitem[Deutsch et al.(2019)]{30}Deutsch A.-S., Dimastrogiovanni E., Fasiello M., Johnson M. C., \& M{\" u}nchmeyer M.\ 2019, Phys. Rev. D, 100, 083538

\bibitem[Dimastrogiovanni et al.(2014)]{24}Dimastrogiovanni E., Fasiello M., Jeong D., \& Kamionkowski M.\ 2014, J. Cosmol. Astropart. Phys., 12, 050




\bibitem[Dodelson et al.(2003)]{15}Dodelson S., Rozo E., \& Stebbins A.\ 2003, Phys. Rev. Lett., 91, 021301

\bibitem[Dodelson(2010)]{16}Dodelson S.\ 2010, Phys. Rev. D, 82, 023522

\bibitem[Emami \& Firouzjahi(2015)]{25}Emami R., \& Firouzjahi H.\ 2015, J. Cosmol. Astropart. Phys., 10, 043

\bibitem[Fabbri \& Pollock(1983)]{8}Fabbri R., \& Pollock M. D.\ 1983, Phys. Lett., 125B, 445

\bibitem[Fassnacht et al.(2006)]{39}Fassnacht C. D., Gal R. R., Lubin L. M., McKean J. P., Squires G. K., \& Readhead A. C. S.\ 2006, ApJ, 642, 30

\bibitem[Frieman et al.(1994)]{33}Frieman J. A., Harari D. D., \& Surpi G. C.\ 1994, Phys. Rev. D, 50, 4895



\bibitem[Grishchuk(1976)]{1}Grishchuk L. P.\ 1976, JETP Lett., 23, 293

\bibitem[Grishchuk(1977)]{2}Grishchuk L. P.\ 1977, Sov. Phys. Usp., 20, 319

\bibitem[Jeong \& Kamionkowski(2012)]{18}Jeong D., \& Kamionkowski M.\ 2012, Phys. Rev. Lett., 108, 251301

\bibitem[Jeong \& Schmidt(2012)]{20}Jeong D., \& Schmidt F.\ 2012, Phys. Rev. D, 86, 083512

\bibitem[Kamionkowski et al.(1997)]{12}Kamionkowski M., Kosowsky A., \& Stebbins A.\ 1997, Phys. Rev. Lett., 78, 2058



\bibitem[Keeton \& Moustakas(2009)]{36}Keeton C. R., \& Moustakas L. A.\ 2009, ApJ, 699, 1720

\bibitem[Linde(1982)]{4}Linde A. D.\ 1982, Phys. Lett. B, 108, 389

\bibitem[Liu(2021a)]{34}Liu W.\ 2021a, Phys. Rev. D, 103, 103012

\bibitem[Liu(2021b)]{35}Liu W.\ 2021b, arXiv:2111.02404

\bibitem[Masui \& Pen(2010)]{17}Masui K. W., \& Pen U.-L.\ 2010, Phys. Rev. Lett., 105, 161302

\bibitem[Rubakov et al.(1982)]{7}Rubakov V. A., Sazhin M.V., \& Veryaskin A.V.\ 1982, Phys. Lett., 115B, 189

\bibitem[Sahni(1990)]{10}Sahni V.\ 1990, Phys. Rev. D, 42, 453

\bibitem[Schmidt \& Jeong(2012)]{19}Schmidt F., \& Jeong D.\ 2012, Phys. Rev. D, 86, 083527


\bibitem[Schmidt \& Jeong(2012)]{21}Schmidt F., \& Jeong D.\ 2012, Phys. Rev. D, 86, 083513


\bibitem[Schmidt et al.(2014)]{23}Schmidt F., Pajer E., \& Zaldarriaga M.\ 2014, Phys. Rev. D, 89, 083507

\bibitem[Seljak(1994)]{37}Seljak U.\ 1994, ApJ, 436, 509

\bibitem[Seljak \& Zaldarriaga(1997)]{13}Seljak U., \& Zaldarriaga M.\ 1997, Phys. Rev. Lett., 78, 2054

\bibitem[Starobinsky (1979)]{9}Starobinsky A. A.\ 1979, JETP Lett., 30, 682



\bibitem[Starobinsky(1980)]{3}Starobinsky A. A.\ 1980, Phys. Lett. B, 91, 99



\bibitem[Starobinskii(1985)]{6}Starobinskii A.\ 1985, Sov. Astron. Lett., 11, 133

\bibitem[Suyu et al.(2017)]{40}Suyu S. H., et al.\ 2017, MNRAS, 468, 2590

\bibitem[Tie \& Kochanek(2018)]{41}Tie S. S., \& Kochanek C. S.\ 2018, MNRAS, 473, 80

\bibitem[Vlah et al.(2020)]{26}Vlah Z., Chisari N., \& Schmidt F.\ 2020, J. Cosmol. Astropart. Phys., 01, 025

\bibitem[Vlah et al.(2021)]{27}Vlah Z., Chisari N., \& Schmidt F.\ 2021, J. Cosmol. Astropart. Phys., 05, 061






































\end{thebibliography}
\end{document}